# Fully Compensated Ferrimagnetic Properties of (Cr,Fe)S Compound with a Pyrrhotite-type Structure

Weida Yin[1,2], Masato Miyakawa[1], Satoshi Semboshi[1,3], Noriharu Yodoshi[4], Akira Masago[5,6], Yosuke Kawahito[7,8], Tetsuya Fukushima[6,9], Hisazumi Akai[8,10] and Rie Y. Umetsu[1,11*]

[1]Institute for Materials Research, Tohoku University, Sendai 980-8577, Japan
[2]Graduate School of Engineering, Tohoku University, Sendai 980-8579, Japan
[3]Faculty of Materials for Energy, Shimane University, Matsue 690-8504, Japan
[4]Graduate School of Engineering, Kyushu University, Fukuoka 819-0395, Japan
[5]Japan Agency for Marine-Earth Science and Technology, Yokohama 236-0001, Japan
[6] Center for Spintronics Research Network (CSRN), Graduate School of Engineering Science, Osaka University, Toyonaka 560-8531, Japan
[7]Japan Agency for Marine-Earth Science and Technology, Yokosuka 237-0061, Japan
[8]Graduate School of Engineering, Osaka University, Suita 565-0871, Japan
[9] Research Center for Computational Design of Advanced Functional Materials, National Institute of Advanced Industrial Science and Technology, Tsukuba 305-8560, Japan
[10]Academeia Co. Ltd., Kashiwa 277-0871, Japan
[11] Center for Science and Innovation in Spintronics, Tohoku University, Sendai 980-8577, Japan

[*]Author to whom correspondence should be addressed: rie.umetsu@tohoku.ac.jp

Abstract

To optimize the processing conditions for the (Cr,Fe)S non-equilibrium phase with a pyrrhotite-type structure, the phase states and magnetic properties of the specimens obtained at various sintering temperatures were investigated. A slightly off-stoichiometric composition of $Cr_{23}Fe_{23}S_{54}$ (approximately $(Cr,Fe)_7S_8$) sintered and quenched from 1323 K indicates a single-phase pyrrhotite-type structure with a layered-NiAs-type structure in which vacancies occupy every two layers ($C12/c1$; the space number is 15). The compound shows fully compensated ferrimagnetic behavior at a magnetization compensated temperature of approximately 200 K. The magnetic behavior exhibits a typical N-type ferrimagnet, as predicted by Néel. From X-ray photoelectron spectroscopy analyses, it is found that the compound is composed of $Fe^{2+}$ and

$Cr^{3+}$. The large magnetic coercivity of 38 kOe at 5 K is also unique and can be applied to spintronic devices. Furthermore, changing the quenching temperature enables control of the degree of order of the vacancies in the interlayer and results in tuning of the magnetization compensated temperature. First-principles calculations show a pseudo-gap located at the Fermi level in the up-spin band, suggesting high spin polarization as well as the NiAs-type structure indicated in the previous our report.

1. **Introduction**

Most half-metals reported thus far are ferromagnets. However, half-metals with alternative magnetic characteristics can lead to the development of new spintronic devices. One such case, proposed by van Leuken and de Groot in 1995 [1], is half-metallic fully compensated ferrimagnets (HM-FCFiMs). Several types of materials have been predicted to be HM-FCFiMs based on the theoretical calculations of perovskite oxides ($La_2M'M''O_6$ : $M'$, $M''$ = transition metal) [2], double perovskites ($LaAVRuO_6$ : $A$ = Ca, Sr, and Ba) [3], and Cr-based full- and half-Heusler alloys ($Cr_2MnZ$ ($Z$ = P, As, Sb, and Bi) [4], $CrMnZ$ ($Z$ = P, As, and Sb) [5]). Akai and Ogura predicted the possible existence of HM-FCFiMs in diluted magnetic semiconductors [6] and pnictide or chalcogenide compounds with NiAs-type crystal structures [7].

Recently, several materials showing potential as HM-FCFiMs have been experimentally reported, such as Mn-based bulk Heusler alloys of Mn-V-Fe-Al ($Mn_{1.5}V_{0.5}FeAl$ [8], $Mn_{1.2}Fe_{1.18}V_{0.62}Al$ [9]), $Mn_2V_{0.5}Co_{0.5}Al$ melt-spun ribbons [10], and $Mn_2Ru_xGa$ thin film [11]. $L2_1$-type crystal cells have been determined that exhibit N- or P-type ferrimagnets, as predicted by Néel [12]. A new material for HM-FCFiMs without Heusler alloys were reported for the first by the authors of the previous paper [13]. The material was a (Cr,Fe)S compound with a NiAs-type structure, and the composition was designed based on the concept that the total number of $d$-electrons per magnetic ion must be equal to five [14].

Many types of compounds have been reported considering the equiatomic compositions of Cr-S, Cr-Se Fe-S, and Fe-Se systems, and superconducting FeSe is a well-known material [15]. Although the crystal structure of FeSe is layered, these materials exhibit a hexagonal NiAs-type structure, which includes vacancies in the equilibrium state at an off-stoichiometric composition and has a long-range stacking structure [16,17]. The phases of $Cr_2S_3$, $Cr_3S_4$, $Cr_5S_6$, and $Cr_7S_8$ are in the Cr-S system, while those of $Fe_7S_8$, $Fe_9S_{10}$, $Fe_{10}S_{11}$, and $Fe_{11}S_{12}$ are in the Fe-S system considering the binary phase diagram [18,19]. A phase of the NiAs-type structure exists at high temperature ranges in both phase diagrams of Cr-S and Fe-S systems around the equiatomic composition, and the phase can exist as solid solute systems from the ternary phase diagram of Cr-Fe-S [20]. Over the last 50 years, $FeCr_2S_4$ chalcogenide spinel compound, where a part of the Fe is substituted by another transition element, have attracted attention for their large magnetoresistance [21], dynamic and static Jahn–Teller effects [22-26], topological anomalous Hall effects [27], multiferroicity [28,29], and magnetostructural transformations [30-32]. This compound is stable at low temperatures. However, in our previous study, a new non-equilibrium phase with a NiAs-type crystal structure was identified based on the concept proposed by Akai *et al.*, which is related to the number of *d*-electrons in this system.

In our previous study, slightly off-stoichiometric (Cr,Fe)S with a NiAs-type crystal structure was successfully obtained via powder solid sintering and quenching at high temperatures [13]. The magnetic data showed that total magnetization compensated at approximately 200 K, and the thermomagnetization curve behaved as an N-type ferrimagnet predicted by Néel, although it shifted slightly to a P-type ferrimagnet under the application of a larger magnetic fields [12]. Among the predicted candidates, (Cr,Fe)S is attractive because it is composed of common elements with high Clarke numbers. Although the composition was optimized over a wide composition range [13], sintering and quenching were not performed. Because a small amount of the secondary phase was included in our previous report, the process used to obtain these materials must be further investigated.

In this study, the quenching temperature dependences of the phase state, crystal structure, and magnetic properties of $Cr_{23}Fe_{23}S_{54}$ were investigated to optimize the processing conditions of the compound. It is found that the crystal structure of the obtained single phase is a pseudo-NiAs-type (pyrrhotite-type) structure with stacking by careful analysis in the lower angle region via X-ray diffraction experiments. X-ray photoelectron spectroscopy analyses also reveal that the materials consist of $Fe^{+2}$ and $Cr^{+3}$.

## 2. Experimental procedures

$Cr_{23}Fe_{23}S_{54}$ was synthesized using powder metallurgy. Raw elemental powders of pure Cr, Fe, and S were mixed and compressed via cold uniaxial pressing of 30 MP at room temperature. The obtained cylindrical compact was encapsulated in a quartz tube under an Ar atmosphere, and the furnace temperature was gradually increased to the expected temperature. After reaching the sintering temperature, the compact was sintered for 1 d and then quenched. The sintering temperatures used were 1023, 1123, 1223, 1323, and 1423 K. The microstructure and composition were examined using scanning electron microscopy (SEM) and inductively coupled plasma (ICP) analysis, respectively. The crystal structure was evaluated using X-ray diffraction experiments with Co-$K\alpha$ radiation. Magnetic measurements were performed using a superconducting interference (SQUID) magnetometer and a vibrating sample magnetometer based on a physical property measurement system (Quantum Design, Ltd.).

The $Cr_{23}Fe_{23}S_{54}$ powders, heat-treated at 1323 and 1423 K, were pestle and pressed into pellets under a pressure of 30 MPa. The pellet-shaped samples were subsequently embedded in resin and polished. X-ray photoelectron spectroscopy (XPS) measurements were carried out using a Shimadzu/KRATOS AXIS-Ultra DLD system to evaluate the valence of the constituent elements of Fe and Cr.

## 3. Theoretical calculations

Theoretical calculations were performed using the first-principles calculation package AkaiKKR, in which the all-electron Korringa–Kohn–Rostoker Green's function method was implemented [33]. Calculations were conducted using atomic sphere approximation, in which overlapping spherically symmetric potentials were employed. In this study, we used the local density approximation with the generalized gradient approximation proposed by Perdew, Burke, and Ernzerhof for the exchange-correlation functional [34]. Furthermore, we employed both non-relativistic and scalar relativistic approximations. However, no significant differences were observed among the results.

The model structure was assumed to be of the pyrrhotite-type, characterized by the space group $C2/c$. This prototype structure contains 14 Fe atoms and 16 S atoms per unit cell. The Fe sites are partially occupied by Cr atoms. This configurational disorder was treated using coherent potential approximation [35,36]. The atomic positions were set consistent with the current experimental structural analysis. Eight spin configurations were considered: (i) ferromagnetic alignment of the magnetic moments for all Fe atoms, and (ii) seven different collinear antiferromagnetic configurations for the Fe atoms. These configurations were uniquely determined by the symmetry of the $C12/c1$ structure. The Fe and Cr spins located on the same type of site were forced to couple antiparallel to each other in all configurations.

## 4. Experimental results

Figures 1 shows SEM images of the microstructures of the specimens. The sintering and water quenching (WQ) temperatures were varied from 1023 to 1423 K (the specimens are denoted as 1023 K WQ). The size of one granule was several hundred microns, and it is composed of particles approximately 20-40 μm in size. Smaller particles of less than 10 μm can also be seen in the specimens obtained from lower quenching temperatures, but these may be attributed to the secondary phase. The number of the small particles decreased with increasing quenching temperature, and no such small particles were observed in the specimens quenched at 1323 and 1423 K. It is clear that a single phase can be obtained by quenching at higher temperatures. By comparing the microstructures of the specimens obtained at 1323 and 1423 K, the outside appearances were not very different, and it seems that the grain size was slightly larger at 1423 K than at 1323 K. The compositions evaluated using ICP are listed in Table 1. The results indicate that all obtained specimens exhibit a nominal composition, and the error bars result from the difficulty in evaluating S using the ICP method.

Figure 2(a) shows XRD patterns measured at room temperature using Co-$K\alpha$ radiation for $Cr_{23}Fe_{23}S_{54}$ compounds, which were sintered and quenched at various temperatures from 1023 to 1423 K. Figure 2(b) shows the simulated XRD patterns of the NiAs-type hexagonal and pyrrhotite-type structures. The pyrrhotite-type structure ($C12/c1$; space number: 15; prototype: $Fe_7S_8$) was a layered NiAs-type structure ($P6_3/mmc$; space number: 194). Most importantly, vacancies exist instead of 1/8 Fe atoms. Therefore, the two theta positions at which the main peaks were observed were the same. These systems differed via the existence of peaks around $2\theta = 20°$, and small peaks were distributed throughout the spectra. Comparing the experimentally obtained XRD patterns and the simulated patterns shown in Fig. 2(b), the extra peaks denoted by crosses (×) for the specimens obtained at quenching temperatures below 1223 K did not correspond to either structure. The peaks of the secondary phase can be attributed to the $FeCr_2S_4$ chalcogenide spinel compound because this phase is stable at low temperatures, as shown in the ternary phase diagram of this system [20]. Here, the lattice parameters in the simulated XRD pattens are assumed to be $a = 0.3450$ and $c = 0.5723$ nm in the NiAs-type structure, and $A = 2a\sqrt{3} = 1.1942$ nm, $B = 2a = 0.6885$ nm, $C = 2c (\sin\beta)^{-1} = 1.2905$, and $\beta = 117.5$ in pyrrhotite. The crystal structures of NiAs- and pyrrhotite-type structures are shown in Fig. 3(a). The pyrrhotite-type structure is layered two times in the directions of $A$ and $C$ axes. Fig. 3(b) shows a simplified pyrrhotite-type structure, in which the layers consisting of S are shown in yellow layer. The positions of the vacancies are readily observed at Fe/Cr layers. Here, it is confirmed that the vacancies are ordered every two layers. In the experimentally obtained XRD patterns, it was found that the peak positions around $2\theta = 20°$ coincide with those of the simulated XRD pattern; however, the intensity differed. To investigate the intensities of the 002, 200, and 111 diffraction peaks, partial disordering of the vacancies was introduced in the XRD simulation. The best fit was obtained as indicated by the lowest patterns colored by light blue shown in Fig. 2(b). Here, the vacancy is assumed to be disordered between the ordinate position and the nearest neighbor in the same layer; the structure is shown in Fig. 3(c). The detailed fitting parameters are provided in the *Supplemental Information*. Furthermore, the intensity of the 002 diffraction peak increased in the specimen obtained at 1423 K WQ. Here, further disordering of the vacancy was introduced, as shown in Fig. 3(d), and the vacancies were distributed in the same layer. Interestingly, changing the quenching temperature resulted in a change in the degree of vacancy ordering. It should be noted that in all simulations, it was assumed that the Cr and Fe atoms were always disordered. This assumption was based on the results obtained from the Mössbauer experiments reported by Sokolovich and Bayukov [37]. They investigated the Mössbauer spectra of the compounds in the CrS–FeS system over a wide range of compositions. The specimens were obtained by rapid cooling and spectral analysis showed that the system was a substitutional solid solution with a random distribution of Cr and Fe atoms.

Figures 4 shows the thermomagnetization (*M-T*) curves measured under magnetic fields of 500 and 5 kOe for each specimen. The *M-T* behavior roughly differed between the specimens obtained at lower- and higher-temperature quenching and is consistent with the XRD results and microstructural observations. For the 1423 K WQ specimen, the *M-T* curves show a completely different behavior between the heating and cooling processes under a 500 Oe magnetic field, whereas those in the heating and cooling processes at 5 kOe were overlaid upon one another. The cross point observed around 90 K contributes to the magnetization-compensated temperature; here, the magnetization of the two sublattices cancel each other. For the specimen obtained at 1323 K WQ, the typical N-type behavior of the *M-T* curve, as predicted by Néel, was observed. The Curie temperature is expected to be slightly higher than 400 K. The N-type behavior and another ferri/ferromagnetic component were mixed in the 1223 K WQ specimen. It is clear that the additional magnetic component originated from the magnetic behavior of $FeCr_2S_4$ spinel compound. The *M-T* behavior of the 1123 and 1023 K WQ specimens show a cusp around 70 K, and the Curie temperature is approximately 170 K. The field cooling effect is observed below 70 K, and these behaviors coincide with the reported magnetic properties of the $FeCr_2S_4$ spinel compound [38,39].

The magnetization (*M-H*) curves of the series of specimens measured at 5 K are shown in Fig. 5. The specimens obtained at 1023 and 1123 K WQ show ferrimagnetic-like behavior, and the magnetization reached 12–14 emu/g at 50 kOe. It has been reported that the magnetic moment of $Cr^{3+}$ and $Fe^{2+}$ is antiferromagnetically coupled and the values of the magnetic moments are 2.9 $\mu_B$ and 4.2 $\mu_B$ at 4.2 K, respectively, providing a total magnetic moment of 1.6 $\mu_B$ per formular unit [40,41]. Although the physical properties of the spinel have attracted much research attention [21-32], spinel is the secondary phase in the present analysis, and it is therefore not discussed in depth. The magnetic component of the spinel decreases with increasing quenching temperature. The specimens obtained at 1323 and 1423 K WQ are single-phase, and the *M-H* curve of the 1423 K WQ specimen showed a linear behavior while the 1323 K WQ sample demonstrated an extremely large coercivity $H_c$. The *M-H* curve for 1323 K WQ does not appear to saturate even at a magnetic field of 50 kOe; therefore, the magnetization was measured up to 90 kOe for the sample.

Figure 6(a) shows the *M-H* curves of the 1323 K WQ specimen measured at 5, 100, 200, and 300 K, and Fig. 6(b) that for entire temperature range from 5 to 300 K. The hysteresis loop measured at 5 K becomes closed at approximately 80 kOe, and $H_c$ reaches 38 kOe. $H_c$ decreases with increasing temperature, is near zero in the vicinity of magnetization compensated temperature $T_{comp}$, and increases above this temperature. Similar tendency was also observed in our previous report [13], however, it is thought that the small $H_c$ is due to the minor loop behavior. It is known that $H_c$ diverges at $T_{comp}$ in ferrimagnetic systems, such as rare earth compounds and oxides [42-47]. To determine $H_c$ near $T_{comp}$, a significantly higher magnetic field is required. Figure 6(c) indicates the temperature dependences of $H_c$ and magnetization at 90 kOe ($M_{90\ kOe}$). Here, $H_c$ is not encountered around $T_{comp}$. In the *M-H* curve at 5 K for the 1423 K WQ specimen (Fig. 5), a linear tendency is observed. This could also be associated with the minor loop in this curve because $T_{comp}$ is approximately 90 K, which is lower than that for the 1323 K WQ specimen. Although the divergence of $H_c$ in the vicinity of $T_{comp}$ was not observed in the present polycrystalline specimen, the symptom can be observed in the single crystal of $(Cr,Fe)_7S_8$ obtained by the chemical vapor transport method, which was successfully obtained by our group. The physical properties of these single crystals will be investigated in a future study. The magnetization obtained at 5 K and 90 kOe was approximately 8 emu/g, which was converted to approximately 0.9 $\mu_B$/f.u. in the pyrrhotite-type structure.

Figure 7 shows the XPS results of Cr, Fe, and S for $Cr_{23}Fe_{23}S_{54}$, which was sintered and quenched from 1323 K (Fig. 7(a)–(c)) and 1423 K (Fig. 7(d)–(f)). In Figs. 7(a) and 7(d), broad peaks associated with Cr $2p_{3/2}$ are observed around the binding energy of 575 eV. Here, this system was analyzed considering the reference spectra provided in previous studies [48,49]. The broad peak results from the sum of five peaks, and the peak positions are 573.64. 574.62,

575.29, 576.10, and 575.7 eV in the Fig. 7(a). Here, the first four peaks corelate to that of $Cr_2S_3$ [48], and the last may result from $C_2O_3$. Therefore, the obtained spectra result from $Cr^{3+}$. The analysis presented in Fig. 7(d) was conducted in the same manner, wherein the results are similar to those obtained for the 1323 K WQ specimen. The numerical results of these analyses are listed in Table 2.

Figures 7(b) and 7(e) show the XPS spectra of Fe2$p$ at 1323 and 1423 K WQ, respectively. These peaks are composed of Fe2$p_{3/2}$ and Fe2$p_{1/2}$ and their satellites. The main peak at approximately 709 eV can be assessed via the spectra of Fe2$p_{3/2}$, in which $FeS_2$ and FeO appear at 707. and 709.04 eV, respectively [50]. This means that the covalent nature of Fe is $Fe^{2+}$ and the species are the same as in the spectrum of the 1423 K WQ specimen. The roles of Cr and Fe are defined via their valence. Figures 7(c) and 7(f) show the XPS spectra for S$p$, exhibiting the species S$p_{3/2}$ and S$p_{1/2}$. A comparison between the two samples obtained at different quenching temperatures revealed that the intensities of the $Cr_2S_3$ and $FeS_2$ peaks were reduced in the 1423 K WQ sample. This suggests that the Fe/Cr environment differed between the two specimens. XRD analyses showed that the vacancies in each of the two layers tended to be disordered in the interlayer at 1423 K WQ, which resulted in a higher quenching temperature. A small difference in the environment causes a balance in the strength of the sublattice magnetization, which leads to different $T_{comp}$ values.

## 5. Calculation results

Our calculations showed that the parallel-spin configuration was the most stable for all Fe atoms. Because Fe and Cr align antiparallelly at the same type of site and each possesses a nonzero magnetic moment, a ferrimagnetic spin configuration was realized. The energy differences among the eight spin configurations were relatively small, on the order of several hundred meV per unit cell. Figures 8(a)– (c) show the partial density of states (DOS) of Cr and Fe and the total DOS of $(Cr,Fe)_7S_8$, respectively, for the most stable state. The positive side of the vertical axis shows the majority-spin states, whereas the negative side represents the minority-spin states. In addition, the horizontal axis indicates the energy relative to the Fermi level. Here, Cr and Fe were assumed to be randomly distributed at each of the three 4$f$ sites and one 4$e$ site. The atomic positions of each element were set to be the same as those of the prototype $Fe_7S_8$ pyrrhotite [51], and the lattice parameters were used to obtain the values from the present XRD refinement ($A$ = 1.1942, $B$ = 0.6885, and $C$ = 1.2905 nm). The values of the magnetic moments of Cr and Fe at each of the four aforementioned sites were –2.91, –2.87, –

2.99, and –2.86 $\mu_B$ and 3.25, 3.31, 3.19, and 3.25 $\mu_B$, and these values are insensitive to the positions. The S elements also exhibited small values, and the total magnetic moment was calculated as 3.54 $\mu_B$/f.u. The calculated magnetic moments for each element are listed in Table 3.

The total DOS shown in Fig. 8(c) demonstrates that a gap appeared exactly at the Fermi level in the majority-spin states where the DOS became zero, indicating a half-metallic state. Therefore, these results suggest that $(Cr,Fe)_7S_8$ is a potential HM-FCFiM system. The spin polarization of $(Cr_{0.5}Fe_{0.5})S$ with a NiAs-type structure was found to be 99.7% in our previous study [13]. A high spin polarization is maintained in the pyrrhotite structure, which is a stacked NiAs-type structure. Both Fe and Cr have high-spin configurations, and the crystal field effect of these components is weaker than that of pyrite, a common Fe-S compound. The well-defined $e_g$-$t_{2g}$ crystal field splitting characteristics of pyrite and similar compounds are absent in this system because of its lower symmetry and more anisotropic ligand environment. This half-metallic state arises because the majority-spin Fe-$d$ orbitals are fully occupied, whereas the majority-spin $d$ orbitals of Cr are largely unoccupied. This half-metallic characteristic results from the ferromagnetic alignment of the local states. If the Fe local moments were aligned antiferromagnetically, the system would form an antiferromagnetic superposition of half-metallic states and the spin-dependent gap in the total DOS would vanish. Because the energy difference between the parallel and antiparallel Fe spin configurations calculated in this study is very small, the stable spin configuration may be altered via the temperature, atomic position fluctuations, or the presence of vacancies. Additionally, because the spins of Fe and Cr are antiparallel, the total magnetic moment is relatively small, and some of these derivatives may include altermagnets with net magnetic moments of zero.

In recent years, interest in spintronics has shifted from ferromagnets to antiferromagnets [52,53]. "Altermagnets" [54-57], also known as the third type of magnetic system following ferromagnets and antiferromagnets, have been the subject of much theoretical and experimental research. This new concept combines the properties of ferromagnets and antiferromagnets. Similar to antiferromagnets, the adjacent magnetic moments are aligned in opposite directions, resulting in zero magnetization throughout the material and no external magnetic field leakage. Furthermore, similar to ferromagnets, the electron energy band structure exhibits a broken time-reversal symmetry. This is because electrons experience different "virtual magnetic fields" depending on their spin orientation, enabling the generation of a spin current with a specific spin bias.

In addition to altermagnets, FCFiMs have gained attention because the total magnetization basically small entire temperature range and especially almost zero around the magnetization compensated temperature [58]. Antiferromagnets are promising materials for applications in tunneling magnetoresistance and other devices because of their small total magnetization, which reduces the interelement interference caused by magnetic field leakage, even when devices are integrated in a small space. FCFiMs are also interesting materials for such applications, especially when they exhibit high spin polarization. Although altermagnets are characterized by spin splitting, which depends on the wavenumber, compensated ferrimagnets exhibit a uniform spin splitting in the band structure that is independent of the wavenumber. Therefore, compensated ferrimagnets are promising systems for realizing half-metallic magnetic materials with low magnetizations. Such materials are likely to gain increased attention in spintronics research.

In this study, our research group successfully synthesized a single-crystal sample of $(Cr,Fe)_7S_8$ with a pyrrhotite-type structure. Experiments are currently being conducted to study the magnetocrystalline anisotropy and transport properties of this system.

## 6. Conclusion

In this study, $(Cr,Fe)_{46}S_{54}$ was fabricated at temperatures ranging from 1023 to 1423 K to optimize the sintering and quenching processing conditions for this system. In addition, single phases were obtained by sintering and quenching from 1323 and 1423 K. The XRD profiles were indexed to pyrrhotite-type structure, which refers. It was found that the degree of vacancy ordering differed depending on the quenching temperature, based on the assumption that the Cr and Fe atoms were randomly occupied. The specimens quenched below 1223 K exhibited the precipitation of $Cr_2FeS_4$ spinel.

The single phase of the pyrrhotite shows the typical behavior of an N-type ferrimagnet in their thermomagnetization curves, showing a fully compensated behavior. However, different magnetization compensation temperatures were observed in the 1323 and 1423 K WQ specimens. In particular, the 1323 K WQ specimen exhibited a large hysteresis in its magnetization curve obtained at 5 K, reaching a magnetic coercivity of 38 kOe. The difference in the magnetization compensated temperatures can be attributed to the slight change in the environment of the Cr and Fe atoms caused by the disordering of the vacancies in the interlayer. XPS analysis showed that the valences of the transition atoms were $Cr^{3+}$ and $Fe^{2+}$, which exhibited distinct roles.

First-principles calculations suggested that a high spin polarization was maintained in the pyrrhotite and NiAs-type structures. (Cr,Fe)S pyrrhotite is a unique magnetic material, acting as a fully compensated ferrimagnet with the potential for realizing a half-metallic electronic state.

**Acknowledgments**

We acknowledge support from the Cooperative Research and Development Center for Advanced Materials (CRDAM) of IMR, in particular the valuable support provided by Ms. Kazuyo Omura, and GP-Spin, Tohoku University. This study was supported by a Grant-in-Aid for Scientific Research (A) (MEXT 22H00287), Japan. Computations were performed using the Earth Simulator supercomputer at the Japan Agency for Marine-Earth Science and Technology (JAMSTEC).

**Conflict of Interest**

The authors declare no conflict of interest.

**Data Availability Statement**

The data that support the findings of this study are available on request from the corresponding author.

References

[1] H. van Leuken and R. A. de Groot, "Half-Metallic Antiferromagnets", Phys. Rev. Lett. 74: 1171 (1995).

[2] W.E. Pickett, "Single Spin Superconductivity", Phys. Rev. Lett. 77: 3185 (1996).

[3] J.H. Park, S.K. Kwon, and B.I. Min, "Half-metallic antiferromagnetic double perovskites: LaAVRuO$_6$ (A=Ca, Sr, and Ba)", Phys. Rev. B 65: 174401 (2002).

[4] I. Galanakis, K. Özdoğan, E. Şaşıoğlu, and B. Aktaş, "Ab initio design of half-metallic fully compensated ferrimagnets: The case of Cr$_2$MnZ (Z=P, As, Sb, and Bi)" Phys. Rev. B 75: 172405 (2007).

[5] E. Sasioglu, "Nonzero macroscopic magnetization in half-metallic antiferromagnets at finite temperatures", Phys. Rev. B 79: 100406(R) (2009).

[6] H. Akai and M. Ogura, "Half-Metallic Diluted Antiferromagnetic Semiconductors", Phys. Rev. Lett. 97: 026401 (2006).

[7] N. H. Long, M. Ogura, and H. Akai," New type of half-metallic antiferromagnet: transition metal pnictides", J. Phys.: Condens. Matter 21: 064241 (2009).

[8] R. Stinshoff, G. H. Fecher, S. Chadov, A. K. Nayak, B. Balke, S. Ouardi, T. Nakamura, and C. Felser, "Half-metallic compensated ferrimagnetism with a tunable compensation point over a wide temperature range in the Mn-Fe-V-Al Heusler system", AIP Adv. 7: 105009 (2017).

[9] M. Ghanathe, A. Kumar, M.D. Mukadam, S.M. Yusuf, "Temperature dependent partially compensated to nearly fully compensated magnetic state in half-metallic full Heusler alloy, Mn$_{1.2}$Fe$_{1.18}$V$_{0.62}$Al", J. Magn. Magn. Mater. 561: 169689 (2022).

[10] P.V. Midhunlal, J. A. Chelvane, D. Prabhu, R. Gopalan, and N. H. Kumar, "Mn$_2$V$_{0.5}$Co$_{0.5}$Z (Z = Ga, Al) Heusler alloys: High T$_C$ compensated P-type ferrimagnetism in arc melted bulk and N-type ferrimagnetism in melt-spun ribbons", J. Magn. Magn. Mater. 489: 165298 (2019).

[11] H. Kurt, K. Rode, P. Stamenov, M. Venkatesan, Y.-C. Lau, E. Fonda, and J. M. D. Coey, "Cubic Mn$_2$Ga Thin Films: Crossing the Spin Gap with Ruthenium", Phys. Rev. Lett. 112: 027201 (2014)

[12] M. L. Néel, "Propriétés magnétiques des ferrites; ferrimagnétisme et antiferromagnétisme", Ann. Phys. 12: 137 (1948).

[13] S. Semboshi, R.Y. Umetsu, Y. Kawahito, and H. Akai, "A new type of half-metallic fully compensated ferrimagnet", Sci. Rep. 12:10687 (2022).

[14] H. Akai and M. Ogura, "Half-metallic antiferromagnets", Japanese Patent, 2008-73917 (P2008-73917), in Japanese (2008).

[15] Fong-Chi Hsu, Jiu-Yong Luo, Kuo-Wei Yeh, Ta-Kun Chen, Tzu-Wen Huang, Phillip M. Wu, Yong-Chi Lee,Yi-Lin Huang, Yan-Yi Chu, Der-Chung Yan, and Maw-Kuen Wu,"Superconductivity in the PbO-type structure α-FeSe", PNAS 105:14262–14264 (2008).

[16] per E. F. Bertaut, "Contribution à l'Etude des Structures Lacunaires: La Pyrrhotine" Acta Cryst. 6: 557 (1953).

[17] M. Tokonami, K. Nishiguchi, and N. Morimoto, "Crystal Structure of a Monoclinic Pyrrhotite ($Fe_7S_8$)", Am. Mineralog. 57: 1066-1080 (1972).

[18] Chromium-Sulfur Binary Phase Diagram (1990 Venkatraman M.), ASM Alloy Phase Diagram Database, P. Villars, editor-in-chief; H. Okamoto and K. Cenzual, section editors

[19] Iron-Sulfur Binary Phase Diagram (2005 Waldner P.) c, ASM Alloy Phase Diagram Database, P. Villars, editor-in-chief; H. Okamoto and K. Cenzual, section editors

[20] Chromium-Iron-Sulfur Ternary, Isothermal Section (1988 Raghavan V.) a, ASM Alloy Phase Diagram Database, P. Villars, editor-in-chief; H. Okamoto and K. Cenzual, section editors

[21] A. P. Ramirez, R. J. Cava, and J. Krajewski, "Colossal magnetoresistance in Cr-based chalcogenide spinels", Nature 386: 156–159 (1997)

[22] L. F. Feiner, "Unified description of the cooperative Jahn Teller effect in $FeCr_2S_4$ and the impurity Jahn-Teller effect in $CoCr_2S_4:Fe^{2+}$", J. Phys. C: Solid State Phys. 15: 1515 (1982).

[23] M. Eibschutz, S. Shtrikman, and Y. Tenenbaum, "Magnetically induced electric field gradient in tetrahedral divalent iron: $FeCr_2S_4$", Phys. Lett. A24: 563-564 (1967).

[24] M.R. Spender and A.H. Morrish, "A low-temperature transition in $FeCr_2S_4$", Solid State Commun. 11: 1417-1421 (1972).

[25] L. Brossard, J. L. Dormann, L. Goldstein, P. Gibart, and P. Renaudin, "Second-order phase transition in $FeCr_2S_4$ investigated by Mössbauer spectroscopy: An example of orbital para-to-ferromagnetism transition", Phys. Rev. B 20: 2933-2944 (1979).

[26] A.M. van Diepen and R. P. van Stapele, "Ordered local distortions in cubic $FeCr_2S_4$", Solid State Commun. 13: 1651-1653 (1973).

[27] K. Ohgushi, S. Miyasaka, and Y. Tokura, "Anisotropic anomalous Hall effect of topological origin in carrier-doped $FeCr_2S_4$", J. Phys. Soc. Jpn. 75: 13710 (2006).

[28] J. Bertinshaw, C. Ulrich, A. Günther, F. Schrettle, M. Wohlauer, S. Krohns, M. Reehuis, A.J. Studer, M. Avdeev, D.V. Quach, J. R. Groza, V. Tsurkan, A. Loidl, and J. Deisenhofer,

"FeCr$_2$S$_4$ in magnetic fields: Possible evidence for a multiferroic ground state", Sci. Rep. 4: 6079 (2014).

[29] L. Lin, H.X. Zhu, X.M. Jiang, K.F. Wang, S. Dong, Z.B. Yan, Z.R. Yang, J.G. Wan, and J.M. Liu, "Coupled ferroelectric polarization and magnetization in spinel FeCr$_2$S$_4$", Sci. Rep. 4: 6530 (2014).

[30] V. Felea, S. Yasin, A. Günther, J. Deisenhofer, H.-A. Krug von Nidda, E.-W. Scheidt, D. V. Quach, J. R. Groza, S. Zherlitsyn, V. Tsurkan, P. Lemmens, J. Wosnitza, and A. Loidl, "Ultrasound study of FeCr$_2$S$_4$ in high magnetic fields", J. Phys: Condens. Matter 48: 486001 (2014).

[31] J. Deisenhofer, F. Mayr, M. Schmidt, A. Loidl, and V. Tsurkan, "Infrared-active phonons in the ferrimagnetic and multiferroic phases of FeCr$_2$S$_4$: Evidence for structural distortions", Phys. Rev. B 100: 144428 (2019).

[32] L. Prodan, S. Yasin, A. Jesche, J. Deisenhofer, H. A. K. von Nidda, F. Mayr, S. Zherlitsyn, J. Wosnitza, A. Loidl, and V. Tsurkan, "Unusual Field-Induced Spin Reorientation in FeCr$_2$S$_4$: Field Tuning of the Jahn-Teller State", Phys. Rev. B 104: L020410 (2021).

[33] M. Ogura and H. Akai, "The full potential Korringa–Kohn–Rostoker method and its application in electric field gradient calculations", J. Phys. Condens. Matter. 17: 5741–5756 (2006).

[34] J. P. Perdew, K. Burke and M. Ernzerhof, "Generalized gradient approximation made simple", Phys. Rev. Lett. 77: 3865-3868 (1996).

[35] H. Akai, Electronic structure Ni-Pd alloys calculated by the self-consistent KKR-CPA method. J. Phys. Soc. Jpn. 51: 468-474 (1982).

[36] H. Akai, "Fast Korringa-Kohn-Rostoker coherent potential approximation and its application to FCC Ni-Fe systems", J. Phys. Condens. Matter 1: 8045-8064 (1989).

[37] V. V. Sokolovich and O. A. Bayukov, "Mossbauer spectra of Fe$_x$Cr$_{1-x}$S solid solutions", Phys. Solid State 49: 1920–1922 (2007).

[38] T. Kanomata, K. Shirakawa, and T. Kaneko, "Effect of Hydrostatic pressure on the Curie temperature of FeCr$_2$S$_4$ and CoCr$_2$S$_4$", J. Phys. Soc. Jpn. 54: 334-338 (1985).

[39] V. Tsurkan, J. Hemberger, M. Klemm, S. Klimm, A. Loidl, S. Horn, and R. Tidecks, "Ac susceptibility studies of ferrimagnetic FeCr$_2$S$_4$ single crystals" J. Appl. Phys. 90: 4639-4644 (2001).

[40] Colominas, C. Broquetas, R. Ballestracci, G. Roult, "Étude pardiffraction neutronique du spinelle FeCr$_2$S$_4$", J. Phys. 25: 526−528 (1964).

[41] G. Shirane, D. E. Cox, and S. J. Pickart, "Magnetic Structures in FeCr$_2$S$_4$ and FeCr$_2$O$_4$", J. Appl. Phys. 35: 954–955 (1964).

[42] T.-H. Wu, H. Fu, R.A. Hajjar, T. Suzuki, M. Mansuripur, "Measurement of magnetic anisotropy constant for magnetooptical recording media: A comparison of several techniques" J. Appl. Phys.73: 1368-1376(1993).

[43] J. Ostorero, M. Escorne, A. Pecheron-Guegan, F. Soulette, H. Le Gall, "$Dy_3Fe_5O_{12}$ garnet thin films grown from sputtering of metallic targets", J. Appl. Phys.75: 6103-6105 (1994).

[44] D.J. Webb, A.F. Marshall, Z. Sun, T.H. Geballe, R.M. White, "Coercivity of a macroscopic ferrimagnet near a compensation point", IEEE Trans. Magn. 24: 588-592 (1988).

[45] S. Demirtas, M.R. Hossu, R.E .Camley, H.C. Mireles, A.R. Koymen, "Tunable magnetic thermal hysteresis in transition metal (Fe, Co, CoNi)/rare earth (Gd) multilayers", Phys. Rev. B72: 184433 (2005).

[46] L.T. Tsymbal, Y.B. Bazaliy, V.N. Derkachenko, V.I. Kamenev, G.N. Kakazei, F.J. Palomares, P.E. Wigen, "Magnetic and structural properties of spin-reorientation transitions in orthoferrites", J. Appl. Phys. 101: 123919 (2007).

[47] E.E. Zubov, "Crystal field and origin of exchange bias in compensated rare-earth ferrimagnets", J. Magn. Magn. Mater. 551: 169153 (2022).

[48] M.C. Biesinger, C. Brown, J.R. Mycroft, R.D. Davidson and N.S. McIntyre, "X-ray photoelectron spectroscopy studies of chromium compounds", Surf. Interface Anal. 36: 1550–1563(2004).

[49] M.C. Biesinger, B.P. Payne, A.P. Grosvenor, L.W.M. Lau, A.R. Gerson, R.St.C. Smart, "Resolving surface chemical states in XPS analysis of first row transition metals, oxides and hydroxides: Cr, Mn, Fe, Co and Ni", Appl. Surf. Sci, 257: 2717–2730 (2011).

[50] M.V. Morales-Gallardo, A.M. Ayala, Mou Pal, M.A. Cortes Jacome, J.A. Toledo Antonio, N.R. Mathews, "Synthesis of pyrite $FeS_2$ nanorods by simple hydrothermal method and its photocatalytic activity", Chem. Phys. Lett. 660: 93–98 (2016).

[51] A.V. Powell, P. Vaqueiro, K.S. Knight, L.C. Chapon, and R.D. Sánchez, "Structure and magnetism in synthetic pyrrhotite $Fe_7S_8$: A powder neutron-diffraction study", Phys. Rev. B 70: 014415 (2004).

[52] J. Železnŷ, P. Wadley, K. Olejník, A. Hoffmann and H. Ohno, "Spin transport and spin torque in antiferromagnetic devices", Nature Phys. 14: 220-228 (2018).

[53] D. Xiong, Y. Jiang, K. Shi, A. Du, Y. Yao, Z. Guo, D. Zhu, K. Cao, S. Peng, W. Cai, D. Zhu, W. Zhao, "Antiferromagnetic spintronics: An overview and outlook", Fundamental Research 2: 522-534 (2022).

[54] L. Šmejkal, J. Sinova, and T. Jungwirth, "Emerging Research Landscape of Altermagnetism", Phys. Rev. X 12: 040501 (2022).

[55] L. Šmejkal, J. Sinova, and T. Jungwirth, "Beyond Conventional Ferromagnetism and Antiferromagnetism: A Phase with Nonrelativistic Spin and Crystal Rotation Symmetry", Phys. Rev. X 12: 031042 (2022).

[56] Y. Liu, S.-D. Guo, Y. Li, and C.-C. Liu, "Two-Dimensional Fully Compensated Ferrimagnetism", Phys. Rev. Lett., 134: 116703 (2025).

[57] N. Cheng, H. Cheng, X. Zhao, G. Hu, X. Yuan, and J. Ren, "Ferroelectric polarization manipulates the layer-polarized anomalous Hall effect in bilayers with fully compensated ferrimagnetism", Phys. Rev. B, 111: 195154 (2025).

[58] T. Kawamura, K. Yoshimi, K. Hashimoto, A. Kobayashi, and T. Misawa, "Compensated Ferrimagnets with Colossal Spin Splitting in Organic Compounds", Phys. Rev. Lett. 132: 156502 (2025).

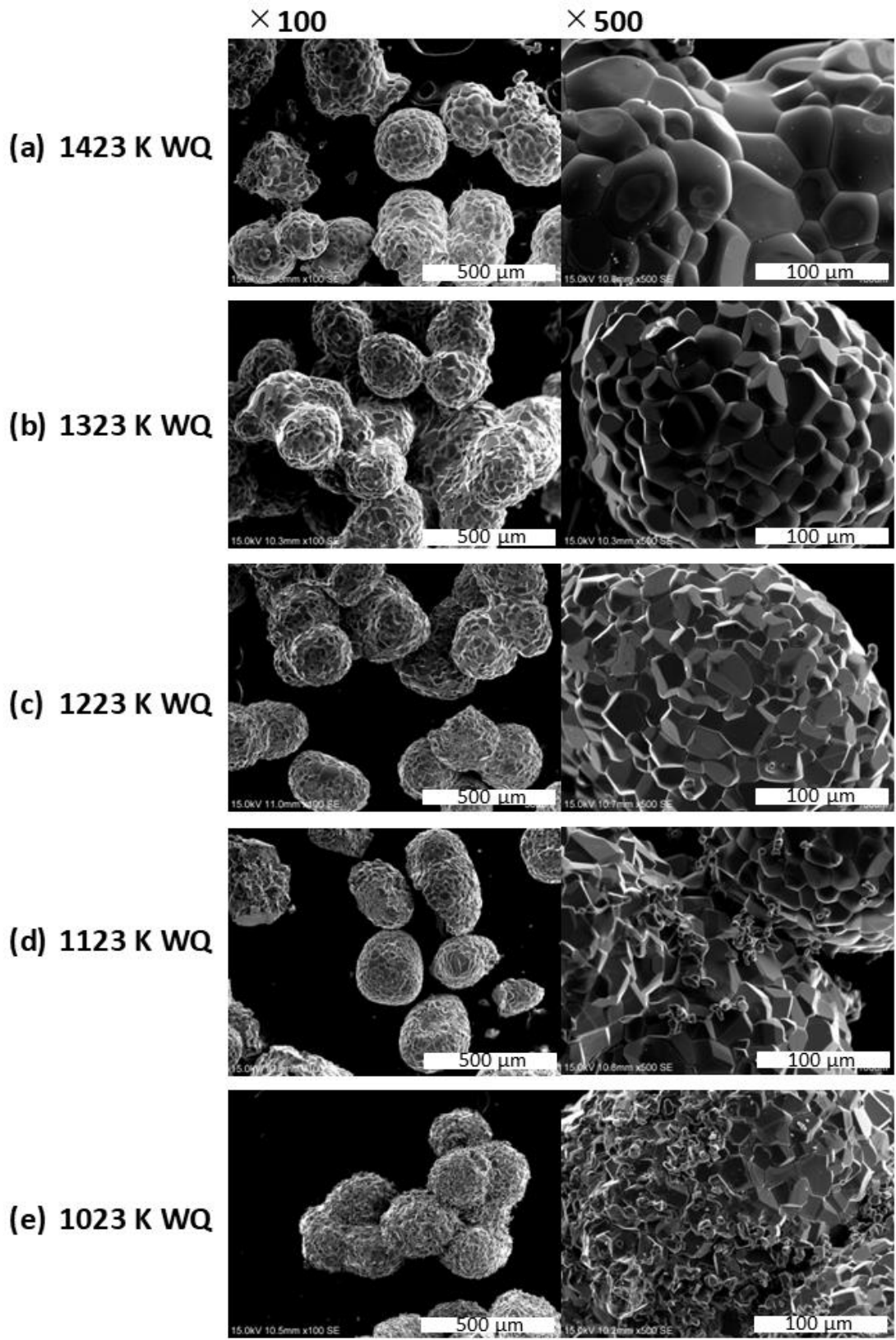

Figure 1. Microstructures observed via SEM for the specimens obtained by sintering and quenching at various temperatures ranging from (a) 1423 K to (e) 1023 K.

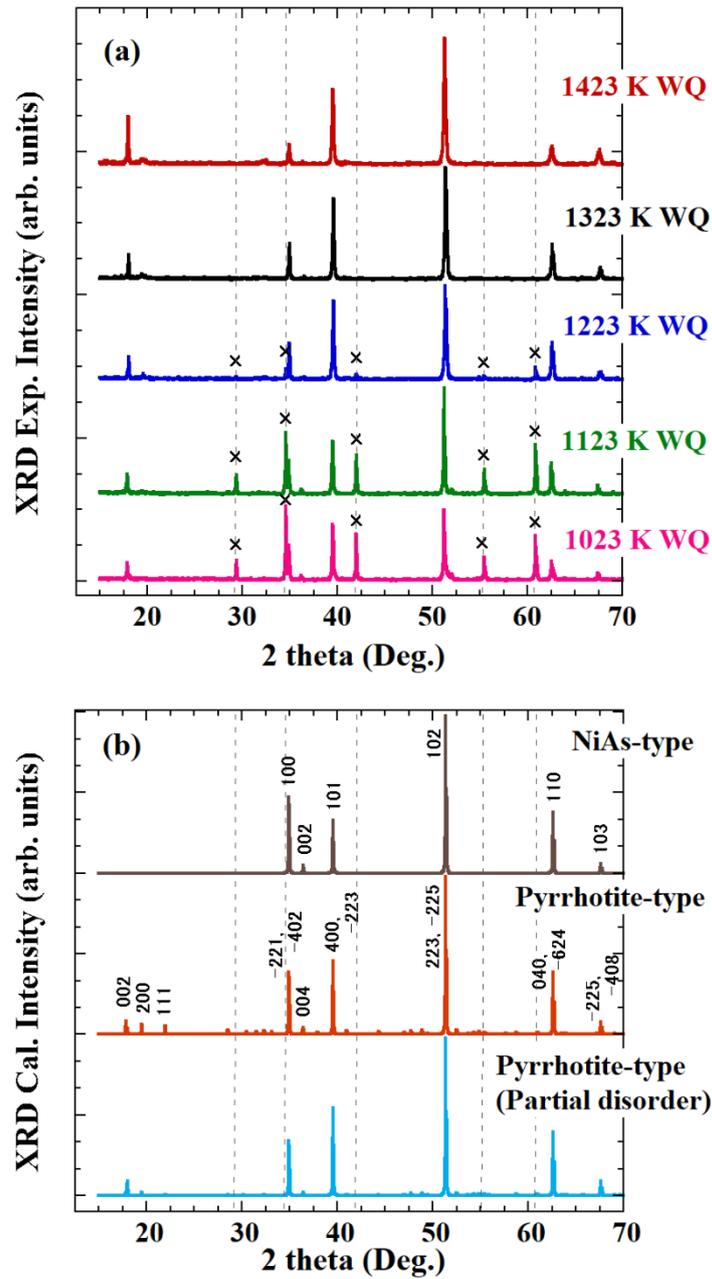

Figure 2. (a) XRD patterns for $Cr_{23}Fe_{23}S_{54}$ compounds sintered and quenched at various temperatures measured using a Co-$K\alpha$ radiation source at room temperature. Peaks indicated with crosses (x) correspond to the secondary phase. (b) Simulated XRD pattens for the (Cr,Fe)S NiAs-type structure and $(Cr,Fe)_7S_8$ with a pyrrhotite-type structure. The middle pattern colored by orange was obtained under the assumption that vacancies order at a certain position (4e site. The lowest pattern colored by light blue was obtained assuming that the vacancies were partially disordered inside of the layer.

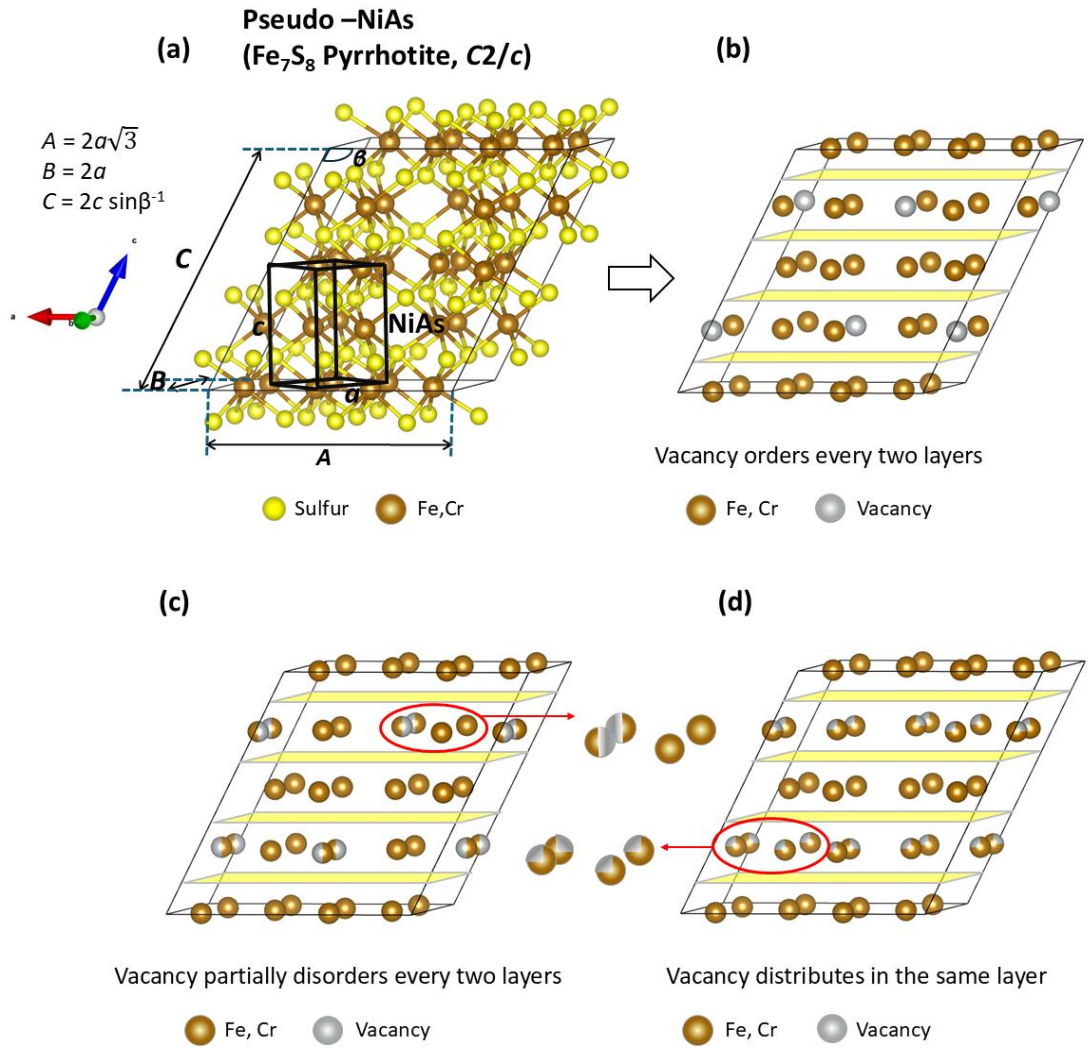

Figure 3. Crystal structure of the prototype pyrrhotite structure (Fe$_7$S$_8$) containing along-range stacking of the NiAs structure, as indicated by the solid line. Here, (a) and (b) show the same structure. Here, the layers constructed of S atoms are indicated as yellow layer simply, in addition, the vacancies are shown by gray in (b). The vacancies are assumed to be disordered inside of each layer in (c) and (d). They partially disorder in (c), but they perfectly disorder inside of the layer in (d).

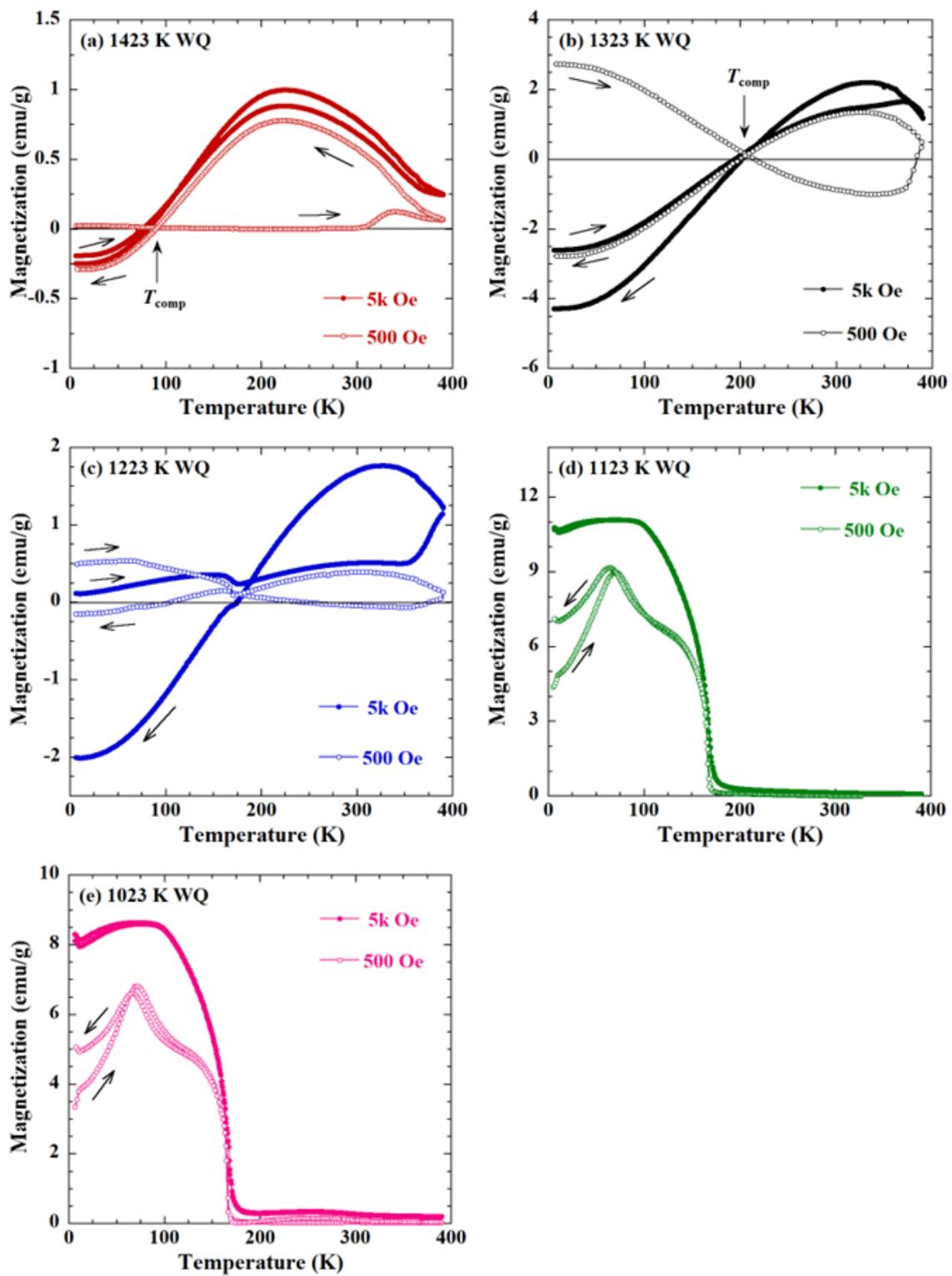

Figure 4. Thermomagnetization curves measured under magnetic fields of 500 and 5 kOe. Figures (a)–(e) correspond to the specimens obtained by quenching at 1423, 1323, 1223, 1123, and 1023 K, respectively.

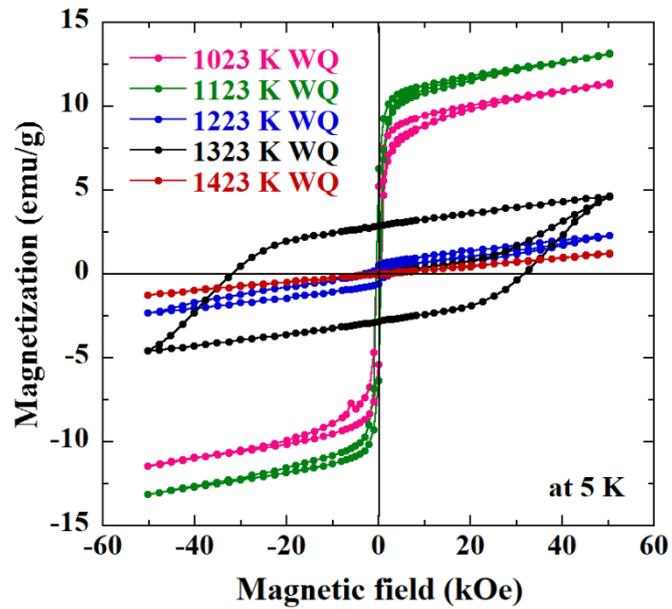

Figure 5. Magnetization curves measured at 5 K up to magnetic field of 50 kOe for each specimen obtained at different quenching temperatures.

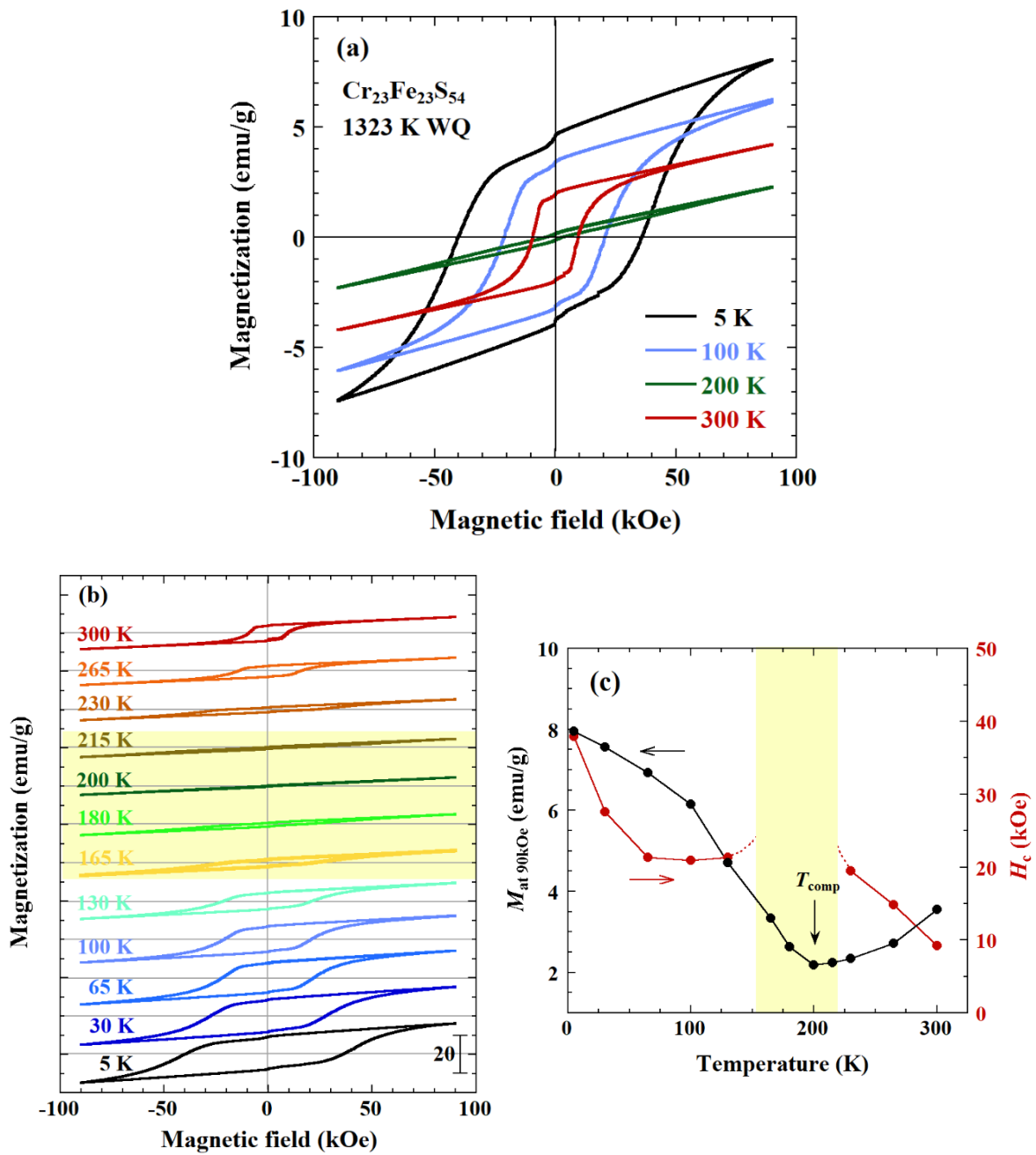

Figure 6. (a) Magnetization (*M-H*) curves measured at 5, 100, 200, and 300 K for the specimen of sintered and quenched at 1323 K. (b) *M-H* curves obtained at temperatures ranging from 5 to 300 K. (c) Temperature dependences of the magnetization obtained at 90 kOe ($M_{90\,kOe}$), and the coercivity ($H_c$). (b) and (c) are also obtained for the specimen of 1323 WQ.

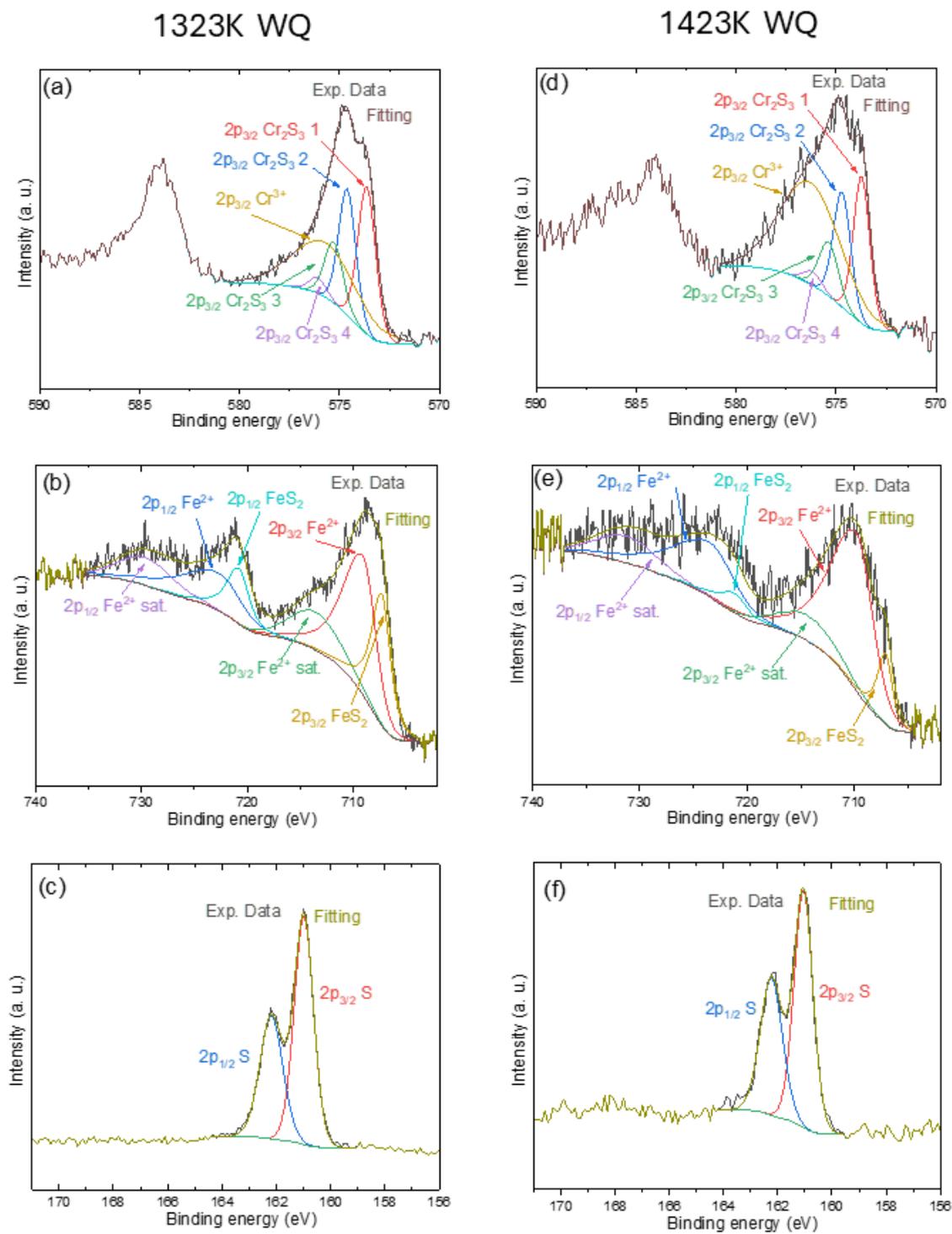

Figure 7. XPS analysis of Cr, Fe, and S in $Cr_{23}Fe_{23}S_{54}$ sintered at (a)–(c) 1323 K and (d)–(f) 1423 K.

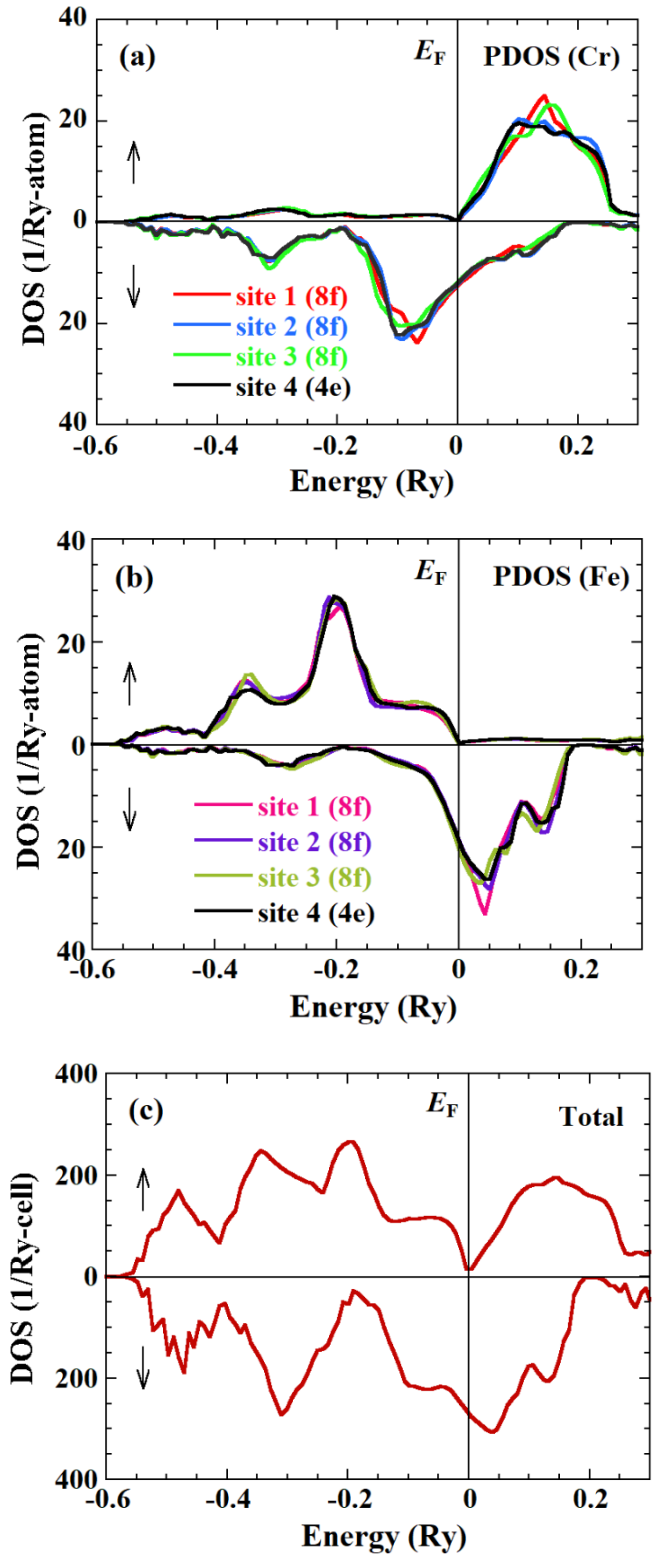

Figure 8. Partial density of states (DOS) of Cr (a) and Fe (b) and the total DOS (c) of $(Cr,Fe)_7S_8$. Here, Cr and Fe are assumed to be randomly distributed on the Fe sites in $Fe_7S_8$ pyrrhotite structure. The atomic positions of each element are set to be the same as those in the prototype $Fe_7S_8$.

Table 1. Compositions of $Cr_{23}Fe_{23}S_{54}$ compounds sintered and quenched from various temperatures evaluated using the ICP method.

| Sintering and quenching temperature (K) | Cr (at.%) | Fe (at.%) | S (at.%) |
|---|---|---|---|
| 1423 | 22.6 | 22.7 | 54.7 |
| 1323 | 24.5 | 21.8 | 53.7 |
| 1223 | 23.9 | 22.4 | 53.7 |
| 1123 | 23.4 | 22.8 | 53.8 |
| 1023 | 24.7 | 21.0 | 54.3 |

Table 2. XPS spectra analysis results. The upper and lower tables corresponded to the specimens obtained at 1323 and 1423 K WQ, respectively.

| | No. | Name | Position | FWHM | Subtotal (%) | Subtotal (%) |
|---|---|---|---|---|---|---|
| 1323 K | | Fe $2p_{3/2}$ 2+ | 709.04 | 2.83 | 18.6 | 26.2 |
| | | Fe $2p_{1/2}$ 2+ | 722.64 | 5.2 | | |
| | | Fe $2p_{3/2}$ 2+ SAT | 713.56 | 5.65 | | |
| | | Fe $2p_{1/2}$ 2+ SAT | 729.68 | 6 | | |
| | | Fe $2p_{3/2}$ $FeS_2$ | 707.29 | 1.98 | 7.6 | |
| | | Fe $2p_{1/2}$ $FeS_2$ | 720.81 | 2.33 | | |
| | 1 | Cr $2p_{3/2}$ $Cr_2S_3$ | 573.64 | 1.05 | 13.7 | 21.7 |
| | 2 | Cr $2p_{3/2}$ $Cr_2S_3$ | 574.62 | 1 | | |
| | 3 | Cr $2p_{3/2}$ $Cr_2S_3$ | 575.29 | 1.09 | | |
| | 4 | Cr $2p_{3/2}$ $Cr_2S_3$ | 576.10 | 1 | | |
| | 5 | Cr $2p_{3/2}$ 3+ | 575.70 | 3.85 | 8.0 | |
| | | S $2p_{3/2}$ Sulfide | 160.98 | 0.84 | 52.1 | 52.1 |
| | | S $2p_{1/2}$ Sulfide | 162.18 | 0.99 | | |

| | No. | Name | Position | FWHM | Subtotal (%) | Subtotal (%) |
|---|---|---|---|---|---|---|
| 1423 K | | Fe $2p_{3/2}$ 2+ | 709.78 | 3.62 | 22.9 | 25.9 |
| | | Fe $2p_{1/2}$ 2+ | 723.38 | 5.24 | | |
| | | Fe $2p_{3/2}$ 2+ SAT | 714.40 | 6.56 | | |
| | | Fe $2p_{1/2}$ 2+ SAT | 731.26 | 6.58 | | |
| | | Fe $2p_{3/2}$ $FeS_2$ | 707.10 | 1.53 | 3.0 | |
| | | Fe $2p_{1/2}$ $FeS_2$ | 721.18 | 2.76 | | |
| | 1 | Cr $2p_{3/2}$ $Cr_2S_3$ | 573.72 | 0.97 | 10.8 | 22.5 |
| | 2 | Cr $2p_{3/2}$ $Cr_2S_3$ | 574.70 | 1.01 | | |
| | 3 | Cr $2p_{3/2}$ $Cr_2S_3$ | 575.37 | 1.13 | | |
| | 4 | Cr $2p_{3/2}$ $Cr_2S_3$ | 576.18 | 1.03 | | |
| | 5 | Cr $2p$ 3+ | 576.30 | 3.56 | 11.7 | |
| | | S $2p_{3/2}$ Sulfide | 161.05 | 0.82 | 51.6 | 51.6 |
| | | S $2p_{1/2}$ Sulfide | 162.23 | 0.99 | | |

Table 3. Magnetic moments obtained via first-principles calculations. The positions of atoms were obtained from the crystallographic information files for $Fe_7S_8$ pyrrhotite ($C12/c1$) [51]. Here, the lattice parameters obtained from the present XRD analysis were used, and Fe and Cr atoms were assumed to be distributed randomly at each Fe site.

|  | Fe ($\mu_B$) | Cr ($\mu_B$) |  | S ($\mu_B$) |
|---|---|---|---|---|
| Site Fe1 (8f) | 3.25 | –2.91 | Site S1 (8f) | 0.10 |
| Site Fe 2 (8f) | 3.31 | –2.87 | Site S2 (8f) | 0.09 |
| Site Fe 3 (8f) | 3.19 | –2.99 | Site S3 (8f) | 0.09 |
| Site Fe 4 (4e) | 3.25 | –2.86 | Site S4 (8f) | 0.12 |

# Supplemental Information

*Model-I*

XRD patterns were further refined by the Rietveld method with Z-Code software [S1,S2]. Two patterns of the models were considered for the specimens obtained by quenching at 1323 and 1423 K for $Cr_{23}Fe_{23}S_{54}$ compound (1323 K WQ and 1423 K WQ, respectively). In the first model for 1323 K WQ, the vacancy is assumed to be distributed in site 4 and site 4*. Fitting results became better when the occupation at the site 2 is not complete 1 rather 0.9. To fix the composition as $(Cr,Fe)_7S_8$, the occupancies at the 4 and site 4* were set to 0.6. Under the assumption, the positions of all atoms were refined. The resultant parameters were listed in Suppl. Table 1 and the fitted patterns were shown in Suppl. Fig. 1. The result is just the one kind of the possibility for the structure model. Further refinement will be required such as neutron diffraction, because Cr and Fe atoms were assumed here to be randomly distributed at the sites 1-4. The lattice parameters were $a = 1.1940$, $b = 0.6887$, $c = 1.2898$ nm and $\beta = 117.5$, the goodness-of-fit indicator $S = 1.4310$.

Suppl. Table 1

| Site | Atoms | Wyckoff | x | y | z | Occ. |
|---|---|---|---|---|---|---|
| Site 1 | Cr0.5 + Fe0.5 | 8f | 0.1337 | 0.1220 | 0.9969 | 1 |
| Site 2 | Cr0.5 + Fe0.5 | 8f | 0.2525 | 0.1213 | 0.2488 | **0.9** |
| Site 3 | Cr0.5 + Fe0.5 | 8f | 0.3678 | 0.1222 | 0.5037 | 1 |
| Site 4 | Cr0.5 + Fe0.5 | 4e | 0 | 0.3973 | 0.25 | **0.6** |
| Site 4* | Cr0.5 + Fe0.5 | 4e | 0 | 0.9025 | 0.25 | **0.6** |
| Site 5 | S | 8f | 0.8989 | 0.1250 | 0.8781 | 1 |
| Site 6 | S | 8f | 0.3537 | 0.1252 | 0.1226 | 1 |
| Site 7 | S | 8f | 0.8519 | 0.1284 | 0.1220 | 1 |
| Site 8 | S | 8f | 0.5994 | 0.1285 | 0.6227 | 1 |

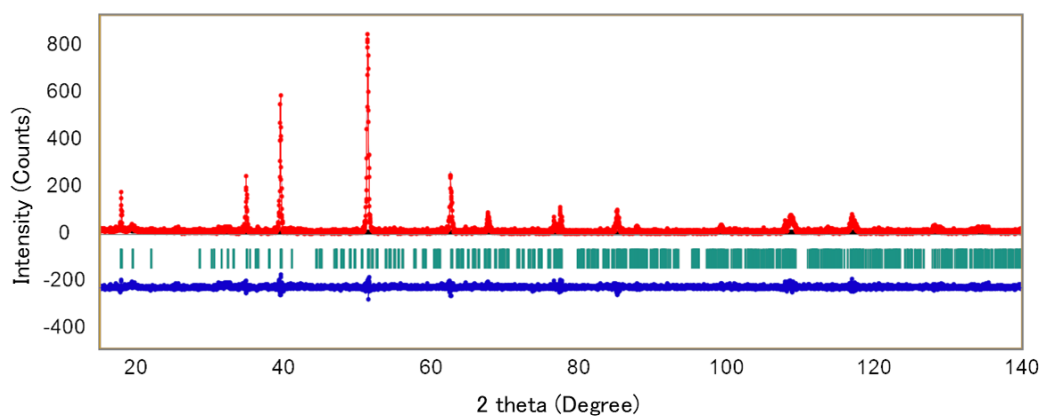

Suppl. Fig. 1 XRD pattern and the Rietveld analysis for the specimen 1323 K WQ.

*Model-II*

This model was used for the refinement for the XRD pattern in the specimen 1423 K WQ. Here, the vacancies were assumed to be distributed at the sites 2, 4 and 4*, therefore, the sits occupancies were set to 0.75. This assumption was the best in the present analyses. Because these sites locate in the same layer, it is natural the disordering may be occurred between them. As in the same manner with the model-I, the atomic positions were refined after fix the site occupancies. The lattice parameters were $a$ = 1.1968, $b$ = 0.6888, $c$ = 1.2938 nm and $\beta$ = 117.5, the goodness-of-fit indicator $S$ = 1.2902.

Suppl. Table 2

|  |  | Wyckoff | x | y | z | Occ. |
|---|---|---|---|---|---|---|
| Site 1 | Cr0.5 + Fe0.5 | 8f | 0.1271 | 0.1170 | 0.9963 | 1 |
| Site 2 | Cr0.5 + Fe0.5 | 8f | 0.2503 | 0.1155 | 0.2496 | **0.75** |
| Site 3 | Cr0.5 + Fe0.5 | 8f | 0.3614 | 0.1172 | 0.5062 | 1 |
| Site 4 | Cr0.5 + Fe0.5 | 4e | 0 | 0.3911 | 0.25 | **0.75** |
| Site 4* | Cr0.5 + Fe0.5 | 4e | 0 | 0.8809 | 0.25 | **0.75** |
| Site 5 | S | 8f | 0.8993 | 0.1277 | 0.8781 | 1 |
| Site 6 | S | 8f | 0.3561 | 0.1332 | 0.1233 | 1 |
| Site 7 | S | 8f | 0.8613 | 0.1325 | 0.1254 | 1 |
| Site 8 | S | 8f | 0.600 | 0.1288 | 0.6206 | 1 |

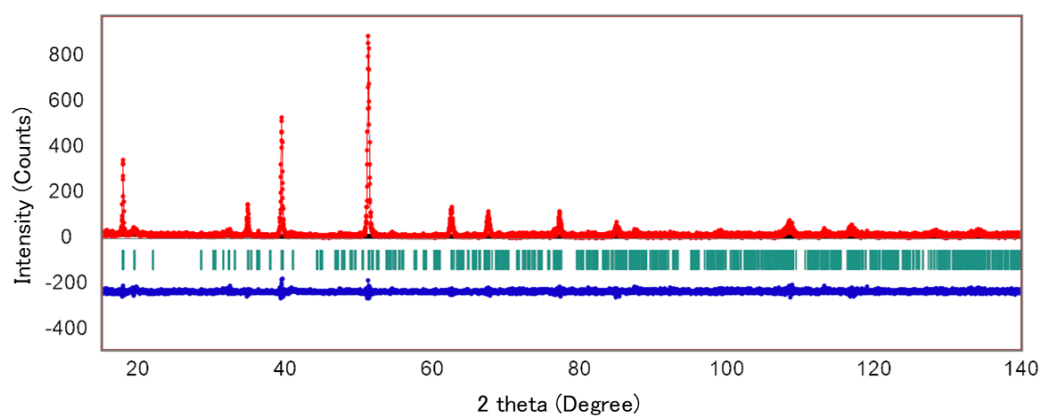

Suppl. Fig. 2 XRD pattern and the Rietveld analysis for the specimen 1423 K WQ.

[S1] R. Oishi, M. Yonemura, Y. Nishimaki, S. Torii, A. Hoshikawa, T. Ishigaki, T. Morishima, K. Mori and T. Kamiyama, "Rietveld analysis software for J-PARC", Nuclear Instruments and Methods A600, 94–96 (2009)

[S2] R. Oishi-Tomiyasu, M. Yonemura, T. Morishima, A. Hoshikawa, S. Torii, T. Ishigaki, T. Kamiyama, "Application of matrix decomposition algorithms for singular matrices to the Pawley method in Z-Rietveld", J. Appl. Cryst. 45, 299–308 (2012)